# Transport anomalies in the layered compound BaPt$_4$Se$_6$

Sheng Li[1], Yichen Zhang[2], Hanlin Wu[1], Huifei Zhai[1], Wenhao Liu[1], Daniel Peirano Petit[1], Ji Seop Oh[2,3], Jonathan Denlinger[4], Gregory T. McCandless[5], Julia Y. Chan[5], Robert J. Birgeneau[3,6], Gang Li[7], Ming Yi[2]*, Bing Lv[1]*

[1]Department of Physics, The University of Texas at Dallas, Richardson, Texas 75080, USA

[2]Department of Physics and Astronomy, Rice University, Houston, Texas 77005, USA

[3]Department of Physics, University of California, Berkeley, Berkeley, California 94720, USA

[4]Advanced Light Source, Lawrence Berkeley National Laboratory, Berkeley, California, 94720, USA

[5]Department of Chemistry, The University of Texas at Dallas, Richardson, Texas 75080, USA

[6]Materials Science Division, Lawrence Berkeley National Laboratory, Berkeley, California, 94720, USA

[7]School of Physical Science and Technology, Shanghai Tech University, Shanghai, 200031 China

* to whom the correspondence should be addressed: mingyi@rice.edu, blv@utdallas.edu

**ABSTRACT**

We report a layered ternary selenide BaPt$_4$Se$_6$ featuring sesqui-selenide Pt$_2$Se$_3$ layers sandwiched by Ba atoms. The Pt$_2$Se$_3$ layers in this compound can be derived from the Dirac-semimetal PtSe$_2$ phase with Se vacancies that form a honeycomb structure. This structure results in a Pt (VI) and Pt (II) mixed-valence compound with both PtSe$_6$ octahedra and PtSe$_4$ square net coordination configurations. Temperature dependent electrical transport measurements suggest two distinct anomalies: a resistivity crossover, mimic to the metal-insulator (M-I) transition at ~150K, and a resistivity plateau at temperatures below 10K. The resistivity crossover is not associated with any structural, magnetic or charge order modulated phase transitions. Magnetoresistivity, Hall and heat capacity measurements concurrently suggest an existing hidden state below 5K in this system. Angle-resolved photoemission spectroscopy measurements reveal a metallic state and no dramatic reconstruction of the electronic structure up to 200K.

**INTRODUCTION**

Two-dimensional (2D) transition metal dichalcogenides (TMDs), with various polytype structures such as 1T, 1T', 2H, and 3R phases, have provided a fertile ground for fundamental quantum materials research and emergent potential applications in the past decade due to the fascinating physical properties discovered in these materials[1–17]. In most cases, structural defects such as vacancies and grain boundaries disrupt the translational symmetry of these pristine lattices and significantly impact their physical and chemical properties. Typically, these defects are detrimental to the carrier mobility and associated charge transport performance[18–20]. However, defect-engineering through carefully controlled atomic defects has recently emerged as a versatile and effective tool which can significantly improve the physical properties, tune the electronic structures, and tailor their device performances for 2D materials. It has been demonstrated that different types of controlled defect structures have led to the enhancement of electrical transport, optical, and chemical properties, for various TMD materials[21–23]. If such defects are well-ordered, a new type of structure or even unprecedented physical properties could also be developed[24,25]. For example, $Mo_6Te_6$ nanowire forms at the boundaries of 2H-$MoTe_2$ through thermal annealing under vacuum[26]; novel $Mo_2S_3$ is fabricated through a periodic assembly of chalcogen vacancy lines in the corresponding $MoS_2$ monolayers[27]. Point defects and line defects of chalcogen atoms induce magnetism in $PtSe_2$ and $ReS_2$ where magnetism does not exist in the pristine materials[28,29]. Besides defect engineering, chemical intercalation that introduces guest species into the *van der Waals* gaps can also effectively change the band filling and the chemical potential of TMD materials and lead to drastic changes of their electronic, transport and optical properties[30–32]. These intercalates can range from cations, anions, neutral atoms and even organic molecules, and could also introduce charge orders[33,34], superconductivity[35,36], or magnetic orders[37,38] into the host TMD materials.

Among the known TMDs, $PtSe_2$ has been well-known for its high performance in photocatalysis, electrocatalysis, and high mobility in field-effect transistors[39–45]. With dimension reduction of $PtSe_2$, the physical properties will change from that of a bulk Dirac semimetal to a monolayer semiconductor with an indirect bandgap of 1.2 eV[46]. Different types of defects have been found in the bulk and thin films of $PtSe_2$, and they have experimentally and theoretically impacted the related properties[28,47]. Thickness-independent semiconducting-to-metallic conversion, in contrast to semiconducting-to-metal transition with reduced thickness, has been observed through plasma-driven atomic defect engineering in $PtSe_2$ films[48]. Theoretical calculations suggested that large spin-orbit splitting can be induced by introducing point defects in the $PtSe_2$[49], and the Se vacancy line defects in monolayer $PtSe_2$ could cause a sizable spin splitting in the defect states[50]. This large spin-orbital splitting gives rise to persistent spin textures which protect the spin from decoherence and induce an extraordinarily long spin lifetime for designing spintronic devices.

As a part of chemical doping studies with controlled defects synthesis for bulk $PtSe_2$ materials[51], we report herein a ternary compound of $BaPt_4Se_6$ of layered structure featured with sesqui-selenide $Pt_2Se_3$ layers sandwiched by Ba atoms. Each sesqui-selenide $Pt_2Se_3$ layer could be considered as a $PtSe_2$ structure with ordered Se vacancies. It has an atomic coordination within the structure of both the PtSe$_6$ octahedra and the PtSe$_4$ square planar, which results in a mixed-valence compound with Pt (II):Pt (IV) = 3:1. It is also interesting to note that the Se atom framework in this compound could be viewed as stacks of distorted Kagome nets formed by Se atoms. We observe a resistivity crossover, mimic to the metal-insulator transition around 150K and a resistivity plateau at temperatures below 10K. The X-ray single crystal diffraction, temperature dependent electrical resistivity, magnetoresistivity, Hall, effect, heat capacity, and angle-resolved photoemission spectroscopy (ARPES) measurements are carried out to fully characterize this compound and understand the intriguing transport anomalies observed in this system.

## RESULTS AND DISCUSSIONS

### X-ray diffraction and structure description

The details of the X-ray single crystal refinements and the associated crystallographic parameters for BaPt$_4$Se$_6$ at both 300K and 100K are provided in Table. 1. Additional atomic coordination, selected interatomic distances, angles, and precession images are presented in the supplementary information. In sharp contrast to PtSe$_2$ in the space group of $P\bar{3}m1$, BaPt$_4$Se$_6$ crystallizes in a distinct structure type (*mC*44) and the centrosymmetric monoclinic space group *C*2/*c* (#15) with lattice parameters *a*=12.715(4) Å, *b*= 7.406 (2) Å, *c*= 12.461 (3) Å, and *β*= 118.089(7) °. The X-ray diffraction pattern with peaks in the preferred orientation is shown in Fig. 1a, together with a photograph of the crystals on the mm scale grid. The crystal structure (Fig. 1b) consists of three distinct Pt sites (Pt1 and Pt2 atoms are at 4*e* sites and Pt3 atoms are at the general position 8*f,* as labeled in Fig. 1b). Pt1 atoms at the 4*e* site are octahedrally coordinated with six neighboring Se atoms and have a formal Pt$^{4+}$ valence. The Pt2 atoms at 4*e* site and the Pt3 atoms at 8*f* site, have a square planar environment with near-by Se atoms, and have a formal Pt$^{2+}$ valence (Supplementary

Figure 1). The PtSe$_6$ octahedra formed by Pt1 atoms are edge shared with the square planar PtSe$_4$ formed by both Pt2 and Pt3 atoms (Fig. 1b). Each PtSe$_6$ octahedra is surrounded by six PtSe$_4$ square planar, and each PtSe$_4$ square planar is shared by two PtSe$_6$ octahedra, together they form a sesqui-selenide Pt$_2$Se$_3$ layer structure with the stacking along the *c*-axis. The interlayer Se-Se distance is 3.478 Å, suggesting weak van der Waals interactions between the sesqui-selenide layers. The Ba atoms are sandwiched between layers and can be charge-balanced as (Ba$^{2+}$)(Pt$^{2+}$)$_3$(Pt$^{4+}$)(Se$^{2-}$)$_6$ with Pt (II):Pt (IV) = 3:1. The Pt (II)-Se distances in the square-planar configuration range from 2.4371(9) Å- 2.4704(9) Å, and are generally shorter than the Pt(IV)-Se distances with octahedral configurations [2.5008(9) Å- 2.5169(9) Å]. Both the PtSe$_6$ octahedra and PtSe$_4$ square planar are highly distorted resulting from the monoclinic symmetry of the crystal structures (Supplementary Figure 1 and Supplementary Table 1 & 2).

The sesqui-selenide Pt$_2$Se$_3$ layer in this BaPt$_4$Se$_6$ compound could be considered as the Se-defect-ordered structure of TMD PtSe$_2$ layers. Fig. 1c represents the projection of the Pt$_2$Se$_3$ slab in BaPt$_4$Se$_6$. In comparison with the projected monolayer PtSe$_2$ slab (Fig. 1d), one-third of the Se atomic positions in the Pt$_2$Se$_3$ slab remain vacant, as highlighted by red open circles, and the vacancies are alternatively shifted to form a well-ordered honeycomb defect pattern. This ordered vacancy pattern is apparently different from the linear vacancy line pattern observed in the Mo$_2$S$_3$ structures and suggests a likely significant change in the electronic structures and transport properties comparing to the parent compound PtSe$_2$.

**Electrical transport data and discussions**

The temperature dependent resistivity data of the BaPt$_4$Se$_6$ are shown in Fig. 2a. In sharp contrast to the overall semi-metallic behavior of the bulk PtSe$_2$, this vacancy-ordered sesqui-selenide shows a drastic resistivity crossover at ~150K with metallic behavior from room temperature down to 150K and semiconducting behavior below 150K. Using the thermal activation model formula of $\rho \propto e^{E_a/k_B T}$ within the temperature range from 150K to 90K, we obtain an activation energy of 0.82 meV. However, the resistivity begins to deviate from the thermal activation model below 80K, starts to saturate below 20K, and reaches a plateau between 10K and 2K, reminiscent of the resistivity plateau observed in SmB$_6$.

Both the resistivity crossover at 150K and resistivity plateau at lower temperatures are quite intriguing, as they hint at a number of potential causes ranging from a magnetic phase transition, charge density wave (CDW) transition, Kondo effect, topological effects, and localization. We therefore carried out temperature dependent magnetization, single crystal X-ray diffraction, and heat capacity measurements to further examine the origins of these transport anomalies. Firstly, no magnetic transition is observed from temperature-dependent magnetization measurement (Supplementary Figure 4). This is consistent with the electron configuration of 5$d^8$ for Pt (II) atoms in the square planar coordination and 5$d^6$ for Pt (IV) atoms in the octahedral coordination as for both configurations the 5$d$ electrons are fully paired. In addition, there are no magnetic ions in this system. Hence resistivity anomalies cannot be induced by the Kondo effect. Secondly, no clear structural transformation is observed from X-ray single crystal diffraction in the low temperature regime down to 100K. The overnight low temperature X-ray single crystal diffraction at 100K is collected and the refined crystal structures are shown in Table 1. Clearly no symmetry changes nor splitting of Wyckoff positions is observed between 100 K and 300 K. The integrated precession images that represent the reciprocal space diffraction spots (Supplementary Figure 2) also do not reflect any CDW-modulated structural distortions at 100K compared with the refined structural model at 300K shown in Table 1. In addition, no clear jumps or anomalies are observed from heat capacity data between 2K and 200K (Fig. 2b), which further supports that no structure transition nor other first order phase transitions exist in this system. Therefore, electron localization is the most likely cause of the resistivity upturn at 150K in this system. A combination of the thermal activation model and variable range hopping model using different dimension indices to represent the strong localization effect have been used to fit the data between 80K and 20K, and yield an unreasonably small activation energy less than 1 μeV. This suggests that the weak localization rather than strong localization exists in the system.

Regarding the origin of the resistivity plateau observed at low temperatures, magnetic effects such as Kondo effect or magnetic field-induced resistivity saturation as observed in WTe$_2$ system are not applicable here. One possible explanation could be the contribution of some additional metallic states, which could originate from multiple conducting channels or surface states competing with the localization effects at low temperatures.

The low temperature range heat capacity data is shown in the inset of Fig. 2b. Interestingly, the heat capacity does not exhibit a linear relationship of $C/T$ vs $T^2$ following the Debye model $C = \gamma_N T + \beta T^3$ between 2K and 10K. By taking the first order derivative of the nonlinear behavior from 2K to 10K, we can clearly observe a broad peak with a maximum at around 6K. This peak may suggest an existing hidden order in the BaPt$_4$Se$_6$ system, which could be responsible for the resistivity plateau observed from the electrical resistivity measurement.

Resistivities with and without applying a magnetic field at 9T in the whole temperature range are shown in Fig. 3a. A clear change of magnetoresistance sign is observed at ~100K, from negative in the low temperature range to positive in the high temperature range. The isothermal magneto resistivity up to 9T at different temperatures are shown in Fig. 3b, which is consistent with Fig. 3a. Three small yet non-negligible anomalies could be observed: (i) a broad hump at 150K which is on par with the M-I transition temperature observed in Fig. 2a; (ii) a change of sign of the magnetoresistance at 100K from positive to negative is observed; (iii) the magnetoresistance upturn below 5K. As the localization are typically induced by disorder, the broad hump at 150K could be explained by the enhanced scattering caused by disorder under magnetic field. However, at low temperatures, a negative magnetoresistivity due to the localization effect will emerge and become stronger with decreasing temperature. The positive and negative magnetoresistance will compensate each other at intermediate temperatures, which could explain the sign changes of magnetoresistance at around 100K. Below 100K, the negative magnetoresistance induced by localization becomes predominant. Interestingly, an upturn of the magnetoresistivity below 5K is observed, which could also be clearly noticed from the inset of Fig. 3b. In comparison with the 5K data, the magnetoresistivity at 2K bends down at high magnetic fields.

In order to demonstrate the weak localization effect in this system, we convert the magnetoresistance data to the magnetoconductance at low field range with temperatures far below 100K to minimize the influence of the classical magnetoresistance and fit the data using the Hikami-Larkin-Nagaoka (HLN) formula[52]

$$\frac{\Delta\sigma(B)}{G_0} = \alpha[\Psi\left(\frac{1}{2} + \frac{B_\phi}{B}\right) - \ln\left(\frac{B_\phi}{B}\right)] \tag{1}$$

which describes the quantum correction to conductivity due to weak localization. In the equation $G_0 = e^2/(2\pi^2\hbar)$, $\alpha$ is the parameter in the renormalization group equation, $\Psi(x)$ is the digamma function, $B_\phi = \hbar/(4eL_\phi^2)$ is the characteristic value of the magnetic field with $L_\phi$ as the phase coherent length. The fitting results are shown in Fig. 3c where the data generally are described reasonably well by the HLN formula, some deviation of the fits from experimental values are observed at our lowest measured temperature of 2K and 5K, which likely is due to the additional metallic states as seen in the previous resistivity data in Fig. 2a. The phase coherence length $L_\phi$ extracted from $B_\phi$, changes from ~47 nm at 2K to ~12 nm at 30K (Fig. 3d). As we increase temperature, the phase coherence length decreases, typically following the power-law relationship $L_\phi \propto T^{-n}$. As the coherence length value at 2K is affected by additional metallic states (Fig. 3c), we therefore only perform a linear fitting from 5K to 30K (red solid line) (Fig. 3d) and obtain a fitted index of $n = 0.62$—a value close to ½ indicating the two-dimensional nature of this system.

In order to further understand the carrier contributions at low temperatures for BaPt$_4$Se$_6$, we performed Hall measurement (Fig. 4a). By sweeping the magnetic field at different temperatures, the Hall resistivity shows a linear relationship with the magnetic field. The Hall resistivity is positive throughout the whole temperature range, suggesting the predominant hole charge carriers in the system. The absolute Hall resistivity value does not change much in the whole temperature range, and the obtained room temperature hole concentration is on the order of $10^{19}$ cm$^{-3}$. This is consistent with the resistivity data where no significant change of resistivity is observed despite the M-I transition at 150K. The temperature dependent Hall resistivity data at 9T is shown in Fig.

4b. Both a broad maximum at 150K and a small upturn below 20K are observed, consistent with both temperature dependent resistivity and magnetoresistivity results discussed previously.

**Band structure calculations and ARPES measurements**

To visualize the electronic structure of BaPt$_4$Se$_6$ and its evolution with temperature, we performed ARPES experiments on the high-quality single crystals. From Density Functional Theory (DFT) calculations on the nominal crystal structure, BaPt$_4$Se$_6$ is expected to be a semiconductor (Fig. 5d). The measured dispersions, however, indicate a metallic state. To introduce the measured dispersions, we first show the three-dimensional (3D) Brillouin zone (BZ) notation of BaPt$_4$Se$_6$ (Fig. 5a) and the corresponding projected (001) plane sketched in grey and appended to the measured Fermi surface (FS) in Fig. 5b. Note that due to the low symmetry of the BaPt$_4$Se$_6$ crystal structure, the FS intensity shows a C$_2$ symmetric pattern with respect to the $\bar{\Gamma} - \bar{M}_2$ direction only. In the $k_z$ mapping displayed in Fig. 5c, the FS shows an anti-symmetric intensity distribution across $k_z$, of which the apparent periodicity is twice that of the BZ size. Such observed intensity pattern is due to the structural factor originating from the two stacked Pt$_4$Se$_6$ layers within each unit cell. Consequently, the $k_z$ FS mapping intensity manifests more strongly the periodicity of the single layer Pt$_4$Se$_6$. The measured FS shows Fermi pockets around the BZ boundaries. Along $\bar{\Gamma} - \bar{M}_2$, we observe these pockets to be small hole pockets from two linear bands crossing the Fermi level consistent with a metallic nature, albeit near the valence band top (Fig. 5e). From a comparison to the calculation, we observe that the bands away from Fermi level present better consistencies with the calculation. The origin of the mismatch here is likely due to Se vacancies in the material that slightly hole-dopes the sample. The electronic dispersions of BaPt$_4$Se$_6$ indicate a degree of resemblance to that of PtSe$_2$, albeit with distinction. The dispersions at the $\bar{\Gamma}$ point emerge from a X-shape feature at near -1.3 eV (Fig. 5f top figure and Supplementary Figure 5). A pair of bands are observed at the $\bar{M}_2$ point, which in PtSe2 are observed to carry spin-texture due to Rashba splitting[45]. However, the measured FS and dispersions show much lower symmetry compared to PtSe2.

We also carried out temperature-dependent measurements to examine the connection to the transport anomalies. The middle panel in Fig. 5f shows the $\bar{\Gamma} - \bar{M}_2$ cut at 151K corresponding to the M-I transition shown above, while the figure at the bottom is taken at temperature high above the transition. Evidently, considering thermal broadening, major features of the band structure, especially the bands crossing Fermi level remain largely unchanged between 14K and 196K. The red triangles in Fig. 5f are fitted from momentum distribution curves (MDCs) to highlight the band crossing the Fermi level. Our observation supports the argument that the resistivity crossover around 150K is not caused by any change of the density of states near $E_F$, but according to the analysis above can be attributed to the weak localization effect.

In summary, A layered ternary compound BaPt4Se6 has been discovered with sesqui-selenide layered feature and mix-valence Pt oxidation states. The sesqui-selenide layer is similar to the TMD PtSe2 structure with ordered Se vacancies. Electrical resistivity measurements reveal two transport anomalies: one resistivity crossover at ~150K and a resistivity plateau below 10K. Extensive studies from X-ray diffraction and heat capacities suggest the resistivity crossover is due to weak localization that is related to the structural distortion and ordered honeycomb Se vacancy of the structure. Temperature-dependent ARPES measurements reveal a metallic state between 14K and 196K, and does not show any abrupt electronic changes in the whole temperature range, further supporting the weak localization effect in this compound. Magnetoresistivity, Hall and heat capacity measurements concurrently suggest an existing hidden state below 5K, which should be responsible for the resistivity plateau at low temperatures. The hole-like band in the MDC curves from the ARPES measurements is consistent with hole charge carriers from our Hall data. Unfortunately, no clear evidence of the existence of the in-gap state is observed from ARPES measurements down to 14K, which is slightly higher than the resistivity plateau below 10K. The exact origin of the emerged metallic ground state at low temperatures responsible for the resistivity plateau is subject for future studies.

**METHODS**

**Material Synthesis and X-ray Diffraction**

The compound was initially discovered during the chemical doping studies for $Ba_xPtSe_2$. The small grain crystals are isolated first for chemical analysis and initial X-ray diffraction studies. The large size of single crystals of $BaPt_4Se_6$ were later directly synthesized through solid state reaction using Ba pieces (99.9%, Alfa Aesar), Pt pieces (99%) and Se shots (99.999%, Alfa Aesar). Stoichiometric amount of the starting materials were placed into a graphite crucible and then sealed in the fused quartz tube under vacuum. The tube was placed in a furnace, slowly heat with a rate of 30°C /h up to 1100 °C, and maintained for 48 hours before slowly cooled down to 800 °C with the rate of 3 °C /h. Thin shinning black crystals can be obtained by carefully cleaving the melt ingot and with preferred *c*-axis orientations (Fig. 1a). The exact crystal structure was determined by Bruker D8 Quest Kappa single-crystal X-ray diffractometer equipped with a Mo Kα IμS microfocus source ($\lambda$ = 0.71073 Å) operating at 50 kV and 1 mA with a HELIOS optic monochromator and a CMOS detector. The collected data set was integrated with Bruker SAINT and scaled with Bruker SADABS (multiscan absorption correction). The refined chemical composition from single crystal diffraction was subsequently confirmed by scanning electron microscope energy-dispersive X-ray spectroscopy (Supplementary Figure 3).

**Physical Properties Measurements**

The electric resistivity and Hall coefficient were measured by employing standard six probe method using golden wires and silver paste on a 2×1×0.1 mm$^3$ crystal with the temperature down to 2K and magnetic field up to 9T. The magnetoresistivity measurement using standard six probe method with magnetic field perpendicular to the crystal plane and current, the heat capacity measurement using the relaxation method down to 2K, were performed in a Quantum Design Physical Property Measurement System. ARPES measurements were performed at the MERLIN beamline 4.0.3 at the Advanced Light Source, equipped with a SCIENTA R8000 electron analyzer. The single crystal was cleaved *in-situ* at 13 K and measured in an ultra-high vacuum with a base pressure better than 5×10$^{-11}$ Torr. Photon energy dependent measurements were carried out from 30 to 120 eV, and 88 eV linear horizontal (LH) light was selected to obtain highest data quality.

**Electronic Structure Calculations**

The electronic structure of BaPt$_4$Se$_6$ was calculated in the framework of DFT within the generalized gradient approximation[53]. The projector-augmented-wave method implemented in Vienna Ab Initio Simulation Package is employed with the default energy cutoff specified in the pseudopotential file. The momentum grid is taken as 9×9×9. The SOC is included in the calculation self-consistently. The calculations are performed at the HPC Platform of ShanghaiTech University Library and Information Services and School of Physical Science and Technology.

## DATA AVAILABILITY

The authors declare that all essential data supporting the findings of this study are available within the paper and its supplementary information.

## ACKNOWLEDGEMENTS

This work at University of Texas at Dallas is supported by US Air Force Office of Scientific Research Grant No. FA9550-19-1-0037. This project is also partially funded by NSF- DMREF-1921581, and the University of Texas at Dallas Office of Research through the Seed Program for Interdisciplinary Research (SPIRe) and Core Facility Voucher Program. J.Y.C. also acknowledges partial support from NSF DMR-1700030. Work at Rice University was supported by the Gordon and Betty Moore Foundation's EPiQS Initiative through grant No. GBMF9470, the Robert A. Welch Foundation Grant No. C-2024, the Alfred P. Sloan Foundation, as well as in part by NSF-DMREF-1921847. Work at UC Berkeley is funded by NSF-DMR-1921798. This research used resources of the Advanced Light Source, a U.S. DOE Office of Science User Facility under contract no. DE-AC02-05CH11231.

## AUTHOR CONTRIBUTIONS

B. L. conceived and initiated this project, S. L., H. W., H. Z., W. L., and D. P. P. performed the sample growth and transport data measurements, Y. Z. and J. S. O. performed the ARPES experiments with the help of J. D. under the guidance of R.J.B. and M. Y., G. T. M. performed the single crystal X-ray diffraction experiments, G. L. performed the theoretical calculations. S. L. M. Y, and B. L. wrote the paper with input from all authors.

## COMPETING INTERESTS

The authors declare no competing interests.

Table 1 Crystal structure of BaPt$_4$Se$_6$ at room temperature and 100K

| Temperature | 300K | | | 100K | | |
|---|---|---|---|---|---|---|
| **Crystal system** | Monoclinic | | | Monoclinic | | |
| **Space group** | C2/*c* (No. 15) | | | C2/*c* (No. 15) | | |
| *a* | 12.715(4) Å | | | 12.709(2) Å | | |
| *b* | 7.406(2) Å | | | 7.4004(14) Å | | |
| *c* | 12.461(3) Å | | | 12.406(2) Å | | |
| *β* | 118.089(7) ° | | | 118.374(8) ° | | |
| **Z** | 4 | | | 4 | | |
| **Atomic position** | | | | | | |
| **Ba1(*4c*)** | 1/4 | 3/4 | 1/2 | 1/4 | 3/4 | 1/2 |
| **Pt1(*4e*)** | 0 | 0.35664(5) | 1/4 | 0 | 0.35783(4) | 1/4 |
| **Pt2(*4e*)** | 0 | 0.85699(5) | 1/4 | 0 | 0.85812(4) | 1/4 |
| **Pt3(*8f*)** | 0.25059(2) | 0.59819(3) | 0.24539(3) | 0.25057(2) | 0.59922(3) | 0.24530(2) |
| **Se1(*8f*)** | 0.03278(6) | 0.61103(8) | 0.13512(7) | 0.03223(5) | 0.61205(7) | 0.13456(5) |
| **Se2(*8f*)** | 0.22316(6) | 0.35380(9) | 0.36289(7) | 0.22376(5) | 0.35499(7) | 0.36357(5) |
| **Se3(*8f*)** | 0.03163(6) | 0.10257(8) | 0.13610(7) | 0.03117(5) | 0.10391(7) | 0.13565(5) |
| **Absorption coefficient** | 78.633 mm$^{-1}$ | | | 79.292 mm$^{-1}$ | | |
| **θ range** | 3.30 to 30.57 ° | | | 3.30 to 30.54 ° | | |
| **Independent reflections** | 1599 | | | 1575 | | |
| **Refine Parameters** | 54 | | | 54 | | |
| **R(int)** | 0.0564 | | | 0.0458 | | |
| **Final R indices** | R$_1$=0.027, wR$_2$=0.058 | | | R$_1$=0.022, wR$_2$=0.055 | | |
| **Goodness-of-fit** | 1.056 | | | 1.076 | | |

**Figure Legends**

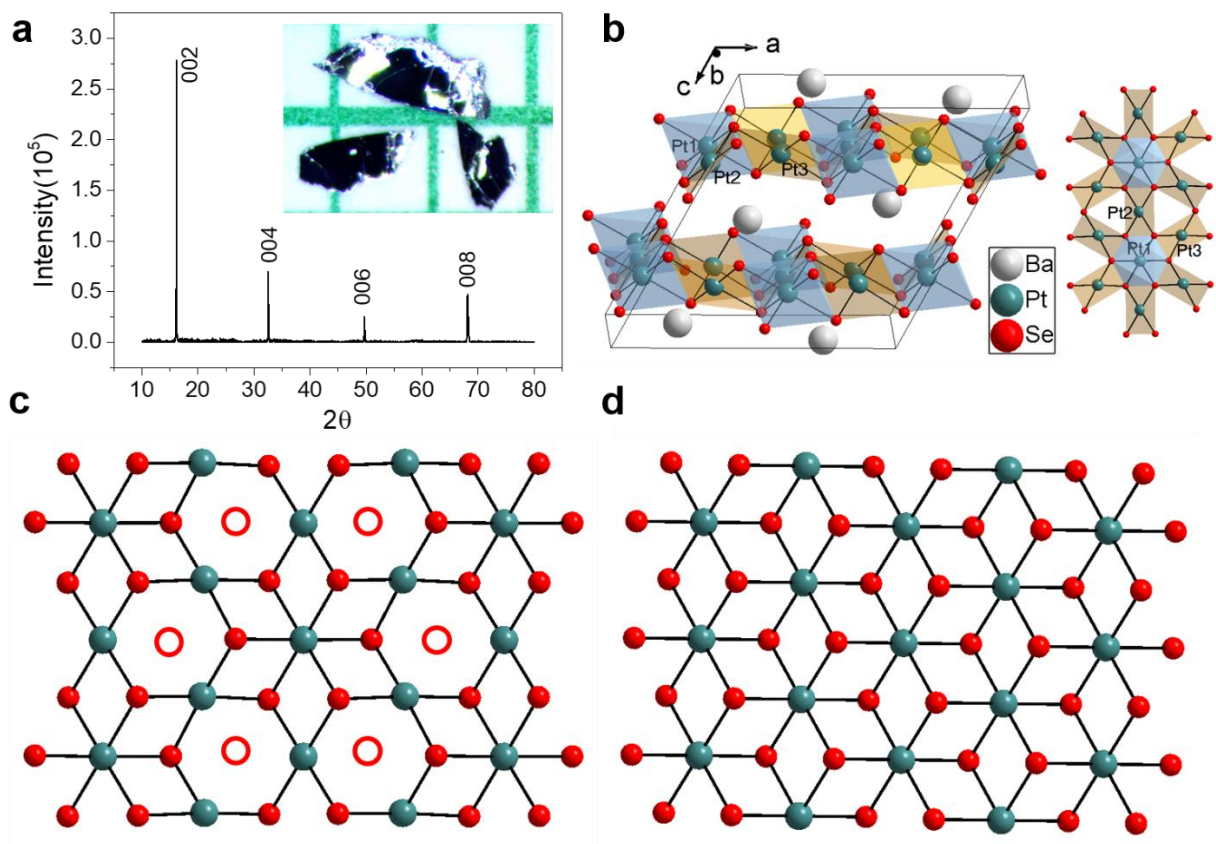

**Figure Legends**

Fig. 1. **Structural analysis of the BaPt$_4$Se$_6$.** **a** Powder X-ray diffraction on single crystal with the preferred orientation along c-axis, inset is the optical image of the BaPt$_4$Se$_6$ single crystals. **b** Side view of BaPt$_4$Se$_6$ structure, with the PtSe$_6$ octahedra and the adjacent PtSe$_4$ square planar. **c** Projection of the Pt$_2$Se$_3$ layer, with the red open circles denote the vacancy of Se atoms and compare to **d** top view of the PtSe$_2$ layer from 1$T$-PtSe$_2$.

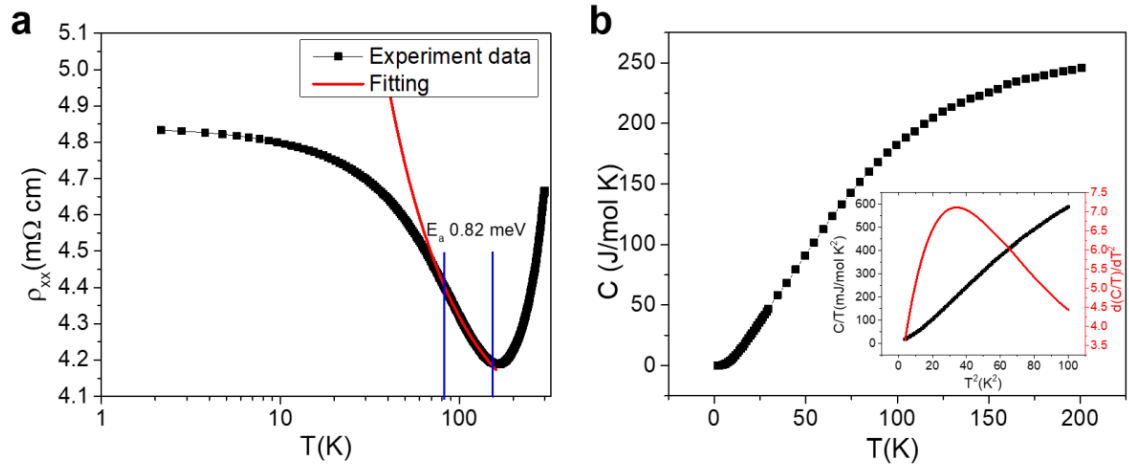

Fig. 2. **Electrical resistivity and heat capacity data. a** Temperature dependent resistivity in log scale to show the low temperature resistivity saturation, the red solid line is the fitted data using the activation energy model. **b** Heat capacity of BaPt$_4$Se$_6$ from 2K to 200K where no clear anomalies or jumps are observed, and the inset is the low temperature heat capacity and first derivative (in red) data from 2K to 10K.

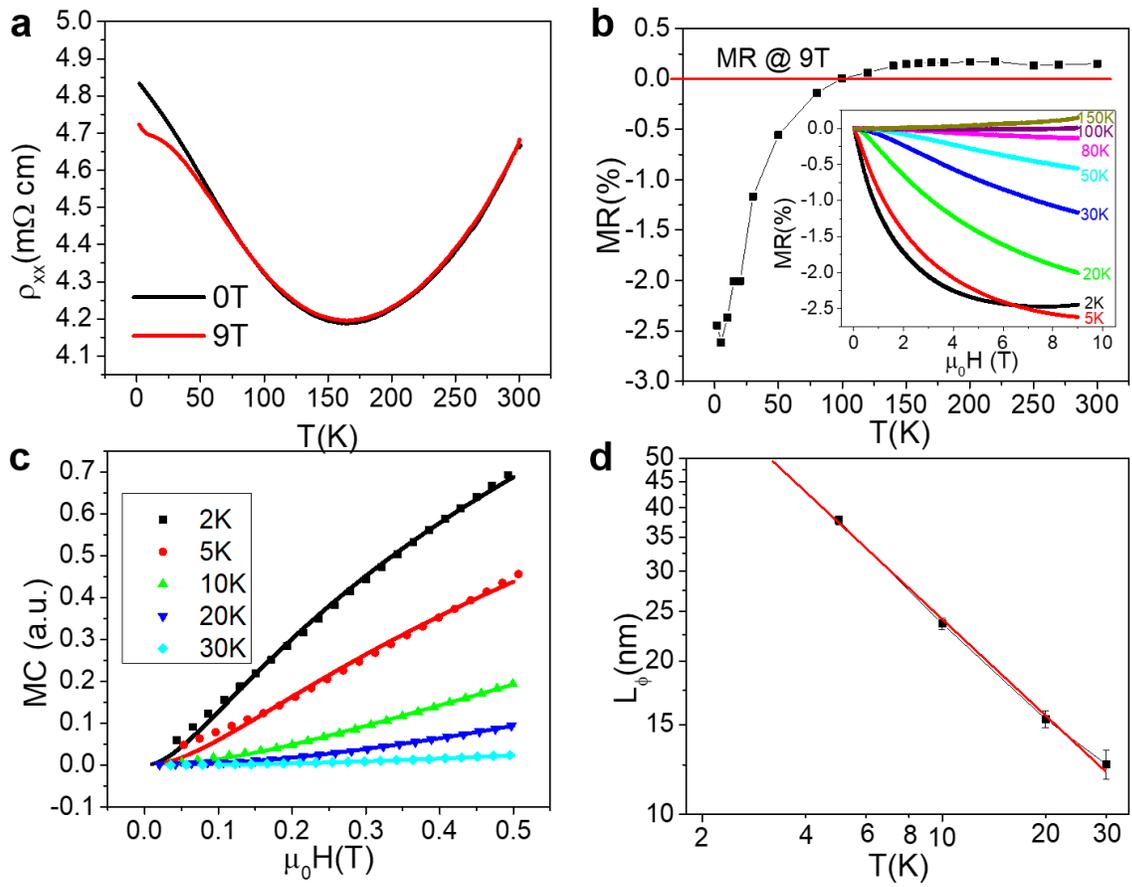

Fig. 3. **In-plane magnetoresistivity data and related analysis**. **a** Temperature-dependent in-plane resistivity data at zero magnetic field and with applied magnetic field up to 9T. **b** Magnetoresistivity at different temperatures under 9T. The inset shows the isothermal magnetic field-dependent resistivity behavior. **c** Magneto conductance at low magnetic field range fitted by the HLN formula (solid line) at several temperatures, and **d** Temperature dependent phase coherence length at different temperatures derived from the HLN formula with linear fitting in the double log scales.

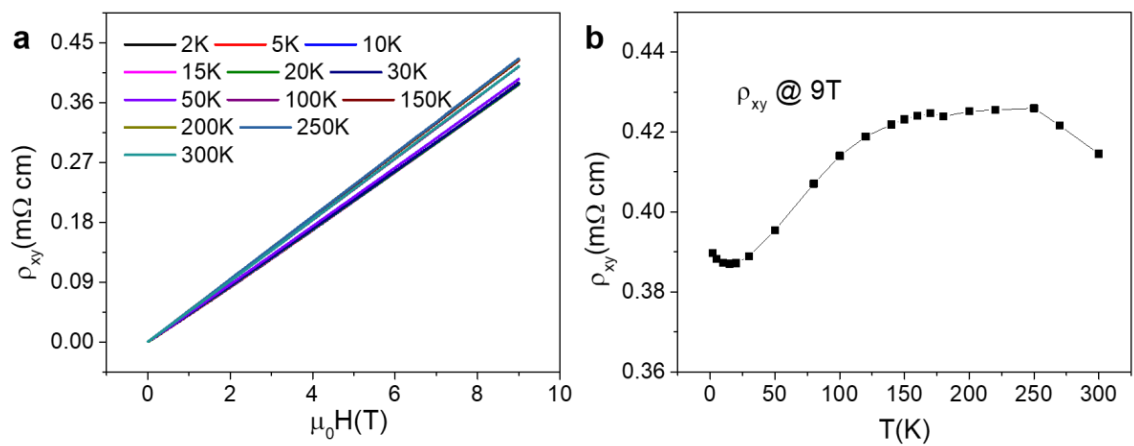

Fig. 4. **Hall resistivity data**. **a** Field-dependent Hall resistivity of BaPt$_4$Se$_6$ at different temperatures up to 9T. **b** Temperature-dependent Hall resistivity at 9T.

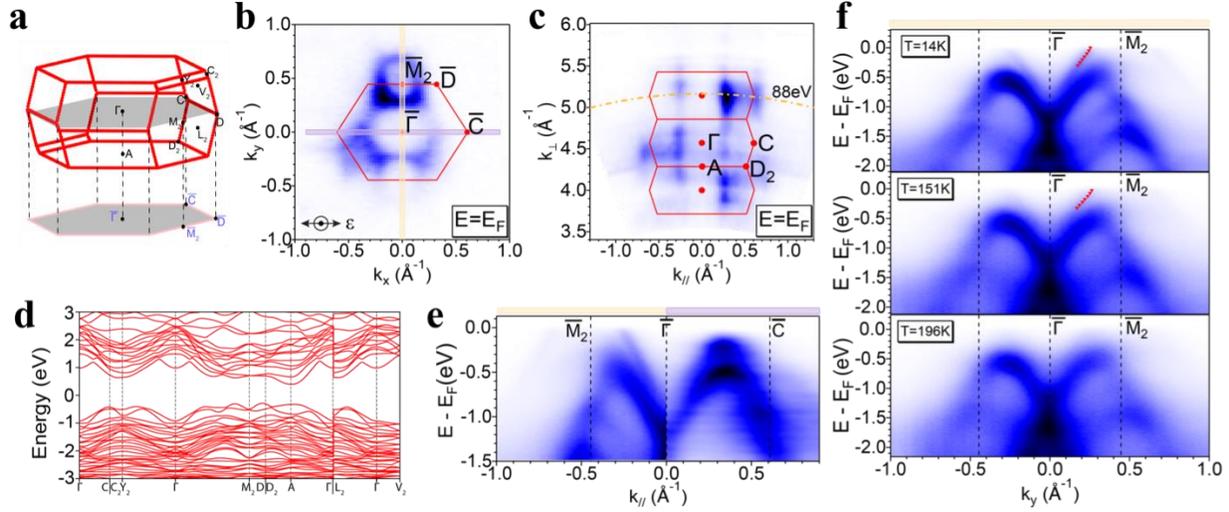

Fig. 5. **Angle-resolved photoemission spectroscopy (ARPES) results**. **a** 3D BZ of BaPt$_4$Se$_6$ where the grey shaded area denotes the projected (001) surface. **b** FS map integrated within a 20 meV window of the projected (001) surface using 88eV LH photons at T=13.5K. Polarization of incident photons is indicated by $\varepsilon$. **c** FS map along the k$_z$ direction. **d** DFT calculation for bulk BaPt$_4$Se$_6$ with SOC included showing a semiconducting ground state. **e** Band dispersions along the $\bar{\Gamma} - \bar{M}_2$ and $\bar{\Gamma} - \bar{C}$ directions where the momentum directions have been marked in **b**. **f** Band dispersions along the $\bar{\Gamma} - \bar{M}_2$ direction at 14K, 151K and 196K, respectively, indicated by a yellow vertical line in panel **b**. Red triangles extracted from MDCs serve as visual guidance for the metallic states crossing the Fermi level.